\documentclass{aa}
\usepackage{graphicx}
\usepackage{psfig}

\newcommand{\gtrsim}{\mathrel{\hbox{\rlap{\lower.55ex \hbox {$\sim$}}
                   \kern-.3em \raise.4ex \hbox{$>$}}}}
\newcommand{\lesssim}{\mathrel{\hbox{\rlap{\lower.55ex \hbox {$\sim$}}
                   \kern-.3em \raise.4ex \hbox{$<$}}}}

\def\ks1731{KS~1731$-$260}

\begin{document}

   \title{A half-a-day long thermonuclear X-ray burst from KS\,1731$-$260}

   \author{E.~Kuulkers
           \inst{1,2}
	   \and
	   J.J.M.~in 't Zand
	   \inst{2,1}
	   \and
	   M.H.~van Kerkwijk
           \inst{2}
	   \and
	   R.~Cornelisse
           \inst{1,2}
	   \and
	   D.A.~Smith
	   \inst{3}
	   \and
	   J.~Heise
           \inst{1}
	   \and
	   A.~Bazzano
           \inst{4}
	   \and
	   M.~Cocchi
	   \inst{4}
	   \and
	   L.~Natalucci
	   \inst{4}
	   \and
	   P.~Ubertini
	   \inst{4}
          }

   \offprints{Erik Kuulkers}

   \institute{
	      SRON National Institute for Space Research,
	      Sorbonnelaan 2, 3584 CA Utrecht, The Netherlands\\
              \email{E.Kuulkers@sron.nl}
	      \and
	      Astronomical Institute, Utrecht University,
	      P.O.\ Box 80000, 3508 TA Utrecht, The Netherlands
              \and
	      University of Michigan, Department of Physics, Ann Arbor, MI, 48109, USA
	      \and
	      Istituto di Astrofisica Spaziale (CNR), Area Ricerca Roma Tor Vergata,
	      Via del Fosso del Cavaliere,
	      I-00133 Roma, Italy
             }

   \date{Received --; accepted --}

   \titlerunning{A superburst from KS\,1731$-$260}

\abstract{
We report on an approximately twelve hour long X-ray flare from the low-mass X-ray binary
\ks1731. The flare has a rise time of less than 13\,min and declines
exponentially with a decay time of 2.7~hours.  The flare emission
is well described by black-body radiation with peak temperature of
2.4\,keV. The total energy release from the event is 10$^{42}$\,erg
(for an assumed distance of 7\,kpc).  The flare has all the
characteristics of thermo-nuclear X-ray bursts (so-called type~I X-ray
bursts), except for its very long duration and therefore large energy
release (factor of 1500--4000 longer and 250--425 more energy
than normal type~I X-ray bursts from this source).  The flare is preceded by a short and
weak X-ray burst, possibly of type~I.  Days to weeks before the flare, type~I X-ray
bursts were seen at a rate of $\sim$3 per day. However, after the 
flare type~I X-ray bursting ceased for at least a
month, suggesting that the superburst affected the type~I bursting behaviour. 
The persistent emission is not significantly different during
the non-bursting period.  We compare
the characteristics of this event with similar long X-ray flares, so-called
superbursts, seen in other sources (4U~1735$-$44, 4U~1820$-$30, 4U~1636$-$53,
Ser~X-1, GX~3+1).  The event seen from \ks1731\ is the longest
reported so far.  We discuss two possible mechanisms that might cause
these superbursts, unstable carbon burning (as proposed recently) and electron
capture by protons with subsequent capture of the resulting neutrons
by heavy nuclei.
  }

   \maketitle

\section{Introduction}

Many low-mass X-ray binaries show thermo-nuclear explosions, or type~I
X-ray bursts (for a review see Lewin et al.\ 1993).  These appear as
rapid ($\sim$1\,s) increases in the X-ray flux, followed by
an exponential decline, with typical durations of the order of seconds
to minutes.  The (net) burst spectra are well described by black-body
emission from a compact object with $\sim$10\,km radius and
temperature of $\sim$1--2\,keV.  The inferred temperature decreases
during the decay, indicating cooling of the neutron star surface.
Typical integrated burst energies are in the 
$\sim$10$^{39}$ to 10$^{40}$\,erg range.

Recently, new modes of unstable burning on neutron stars may have
been revealed, with the discovery of X-ray flares lasting several hours 
(Cornelisse et al.\ 2000; Strohmayer 2000; Strohmayer \&\ Brown
2001; Wijnands 2001; Kuulkers 2001; Cornelisse et al.\ 2001).  The
recurrence times of these events are not well constrained.
Wijnands (2001) reported two events from 4U~1636$-$53 
which were $\simeq$4.7~years apart. So far, only seven events have been 
found, indicating that they are rare.
The availability of X-ray instruments 
such as the BeppoSAX/WFC and the RXTE/ASM, which 
more frequently monitor the X-ray sky, and the RXTE/PCA, which performs
long studies of the X-ray burster population, is
the reason that these events have now been discovered. 

These long X-ray flares share many of the characteristics of type~I bursts,
which is the reason they are attributed to thermonuclear runaway
events (see e.g.\ Cornelisse et al.\ 2000).  What distinguishes them
from normal type~I bursts are the long duration (exponential decay
times of a few hours), the large fluences ($\sim$10$^{42}$\,erg),
and the aforementioned extreme rarity.  Because of the large fluences,
they are also referred to as `superbursts'
So far, these events have only been observed in
sources with persistent pre-burst luminosities of $\sim$0.1--0.3 times
the Eddington luminosity (e.g.\ Wijnands 2001).

The current view is that the superbursts are caused by the unstable
burning of the ashes of the (un)stable hydrogen and/or helium burning
(Cumming \&\ Bildsten 2001; Strohmayer \&\ Brown 2001). Such bursts
in principle thus not only tell us about properties of material buried below the
hydrogen and/or helium layer, but also about the burning of the hydrogen and/or helium layer
itself.

We here report on a superburst seen with the BeppoSAX Wide Field
Camera (WFC) from KS~1731$-$260 on 1996 September 23. 
In Sect.~2, we briefly review what is known
about KS~1731$-$260.  Next, we describe the observations (Sect.~3)
and the results (Sect.~4).  We conclude with a discussion of possible
mechanisms for the superbursts. 
Hereafter we use the term (X-ray) bursts, whenever we refer to type~I (X-ray) bursts
(unless it is needed for clarity).

\section{KS~1731$-$260}

\ks1731\ was discovered as a transient and bursting source in 1989 by 
COMIS/TTM onboard Mir-Kvant (Sunyaev et al.\ 1989, 1990).  
A possible infrared counterpart was identified recently using a precise localisation 
by Chandra (Wijnands et al.\ 2001b; Revnivtsev \&\ Sunyaev 2002; 
Groot et al.\ 2001, in preparation). The source is an X-ray burster
(Sunyaev et al.\ 1990; Muno et al.\ 2000), which identifies the
compact object as a neutron star.  Highly coherent X-ray
oscillations were discovered by the RXTE/PCA at a period of 1.9\,ms 
during several X-ray bursts (Smith et al.\ 1997; Muno et al.\ 2000) as
well as kHz quasi-periodic oscillations
at $\sim$900\,Hz and $\sim$1200\,Hz in the persistent emission
(Wijnands \&\ van der Klis 1997).  The
distance to the source, as derived from the peak luminosity observed
in radius-expansion bursts, is about 7\,kpc (Muno et al.\ 2000).

Fig.~\ref{plot_asm_wfc}a shows the RXTE/ASM light curve of \ks1731\
from near the start of the RXTE mission, i.e.\ 1996 Jan 6, to 2001 May 1. 
The source is highly
variable on time scales of a day to weeks.  During the first two and a
half year of the RXTE mission the overall source intensity varied
slowly on time scales longer than a week.  Subsequently, it showed a
two and a half year interval with strong flares or outbursts on time
scales of weeks to a month, reflecting a change in source state 
(see also Muno et al.\ 2000; Revnivtsev \&\ Sunyaev 2001). 
At the beginning of 2001, the source became quiescent (see Wijnands et al.\ 2001a).

\section{Observations}

BeppoSAX was launched on 1996 April 30. Onboard are two Wide Field
Camera's (WFCs); they consist of coded mask aperture cameras (Jager et al.\
1997). They point in opposite directions of one another and perpendicular to the
Narrow-Field Instruments on the same satellite (Boella et
al. 1997). The field of view of each WFC is 40$\times$40 square
degrees full-width to zero response with an angular resolution of
about 5 arc minutes in each direction.  The band pass of the
instruments is 2 to 28\,keV.  The imaging capability and sensitivity
allow an accurate monitoring of complex sky regions, like the Galactic
Centre.  Two times per year, in spring and autumn, for about two
months a dedicated program is carried out to monitor the Galactic
Centre region typically for about half a day on a weekly basis.  
During the performance verification 
phase, however, the Galactic Centre was monitored continuously (only
interrupted by South Atlantic Anomaly passages and earth occultations) for an interval of
$\sim$9~days, four months after launch (see e.g.\ Cornelisse et al.\
2000; Kuulkers et al.\ 2000). \ks1731\ lies close to the
Galactic Centre ($l=1.07\degr$, $b=+3.66\degr$), ensuring the
source is well covered. In this paper we use the data from 
from either WFC unit 1 or 2, from the
Galactic Centre monitoring program as well as from serendipitous
observations close to the Galactic Centre, obtained in 1996
(net observing time of 851\,ksec), 1997 (785\,ksec) and
northern spring 1998 (414\,ksec).
Data covering (part of) the Galactic Centre are routinely checked for
X-ray bursts. We found a total of 63 bursts from \ks1731\ in
the above interval.

The Rossi X-ray Timing Explorer (RXTE) was launched 1995 December
30. This satellite carries three instruments, including the All-Sky Monitor
(ASM), which consists of three Scanning Shadow Cameras (SSCs) mounted on a
rotating drive such that one is perpendicular to the other two (Levine et
al.\ 1996).  The cameras are held stationary for 90-s intervals, called
`dwells', during which data are accumulated.  The mount rotates the assembly
through a 6$\degr$ angle between dwells until a full rewind is necessary.  In
this manner, $\sim$80\%\ of the sky is observed every 90-min orbit.  Each dwell
provides intensities in the 1.5--12~keV band for all known sources in the field
of view of each camera, and results that meet a set of reliability criteria are
saved in electronic tables that are available over the world wide 
web\footnote{These quick-look results can be found at {\tt http://xte.mit.edu},
and final ASM data products can be found at 
{\tt http://heasarc.gsfc.nasa.gov/docs/xte/asm\_products.html}.}.  
An alternate, time-series, data mode records the total counts in each SSC in 0.125-s bins,
but does not preserve the position of each photon (Levine et al.\ 1996, Smith et
al.\ 2001).  Over the $\sim$5.5 years of RXTE operation, KS~1731--260 was
observed an average of $\sim$10 times a day, yielding $\sim$17000 individual
intensity measurements.

\section{Analysis and results}

\subsection{A long X-ray flare}

\begin{figure}
\resizebox{\hsize}{!}{\includegraphics[clip, bb=34 35 456 732]{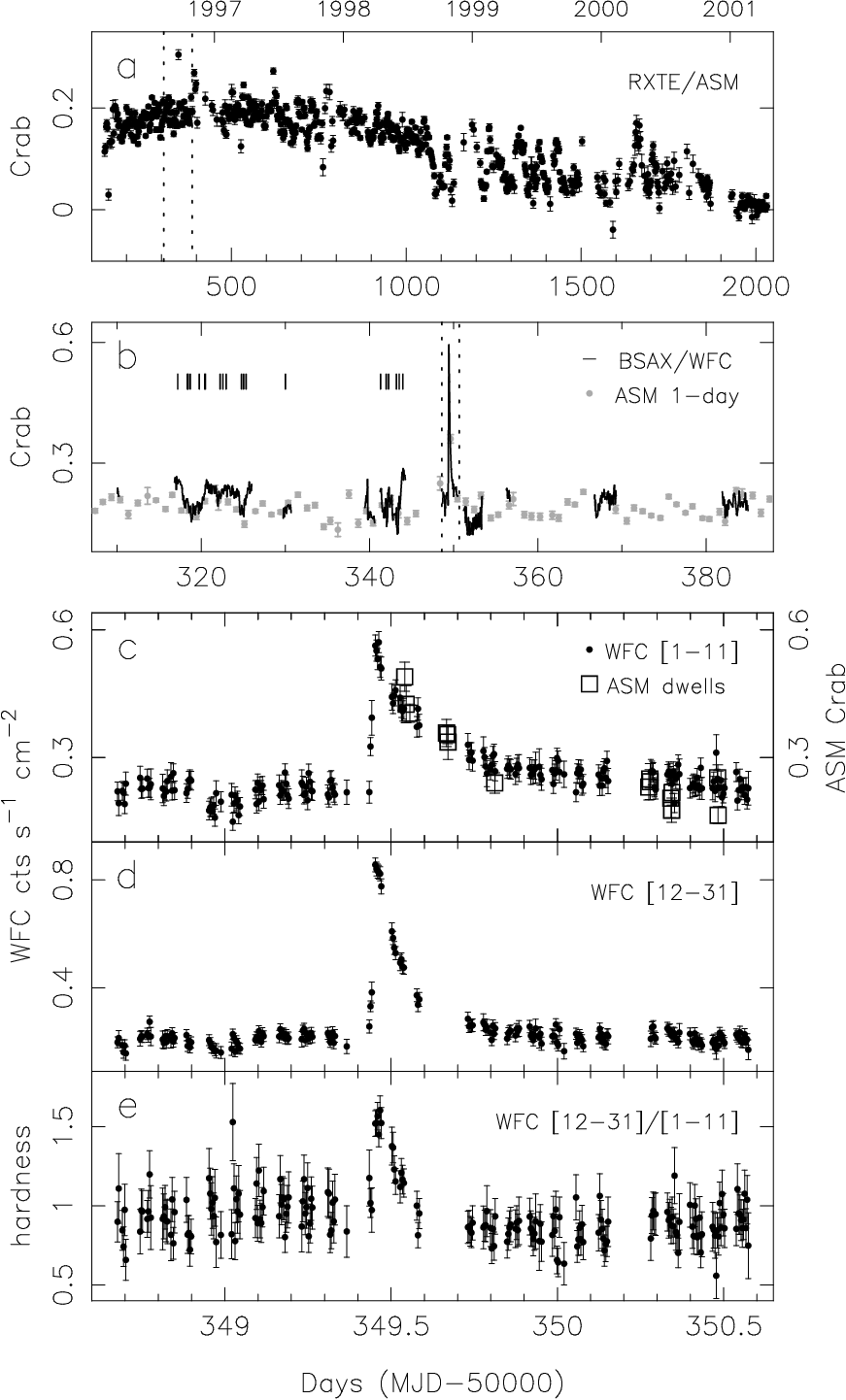}}
\caption*{
{\bf a.} RXTE/ASM (1.5--12\,keV) light curve of \ks1731\ from 
1996 January 6 (MJD\,50088) to 2001 May 1 (MJD\,52030). The data shown
are two-day averages and normalized to the count rate of the Crab
(75\,cts\,s$^{-1}$\,SSC$^{-1}$); only intervals with more than one dwell
per day were included.
{\bf b.} The BeppoSAX/WFC (2--28\,keV) data
during the interval 1996 August 12 to November 1 (MJD\,50307--50388;
marked by the dotted lines in panel {\bf a}). Data points (typically
averages per satellite orbit) which were less than half a day apart
are connected to guide the eye, and are normalized to the Crab
(2\,cts\,s$^{-1}$\,cm$^{-2}$). The gray data points are the one-day averages 
(with more than one dwell per day) of the RXTE/ASM in the same interval.  
A long X-ray flare in the WFC data is seen just before MJD\,50350 
(1996 September 24). The vertical tickmarks indicate the times of the 20 X-ray 
bursts seen with the WFC. They all occur before the flare. 
{\bf c--e.} WFC light curves around the flare (marked by dotted lines
in panel~{\bf b}) for channels 1--11 (2--5\,keV; {\bf c}) and 12--31
(5--28\,keV; {\bf d}), and the hardness ratio of the two above defined
energy bands ({\bf e}). The time resolution is 5\,min. Overplotted in 
{\bf c} with open squares are the data from individual RXTE/ASM dwells.
}
\label{plot_asm_wfc}
\end{figure}

Following the discovery of the long X-ray flare in 4U~1735$-$44
(Cornelisse et al.\ 2000), we found two other such events in our 
WFC database of X-ray bursters:
one from Ser~X-1, presented by Cornelisse et al.\ (2001), the other
from KS~1731$-$260, presented here.  The long X-ray flare of KS\,1731$-$260 can be seen
clearly in Fig.~\ref{plot_asm_wfc}b, which shows the WFC data obtained
on \ks1731\ in 1996, together with the one-day averages of the ASM. A
blow-up of this light curve in two energy bands near the flare
is shown in Figs.~\ref{plot_asm_wfc}c and d.  The flare started
near MJD\,50349.42 (1996 September 23.42 UT), the observed maximum was
seen about half an hour later. Subsequently the source decayed back
exponentially to its pre-flare level with a decay time of $\tau_{\rm
exp}\simeq2.7$\,hr (2--28\,keV; see Table~1).  The decay is faster at
high energies than at low energies (Table~1).  The distinction in the
two decay light curves is shown clearly in the hardness curve
(Fig.~\ref{plot_asm_wfc}e): the spectra harden during the rise to
maximum, and soften again during the subsequent decay.  

The flare was partly covered by the RXTE/ASM
(Fig.~\ref{plot_asm_wfc}c) and corresponds to the highest data point
in the ASM light curve (Fig.~\ref{plot_asm_wfc}a).  This enhancement
was already noticed by Wijnands (2001), who, however, could not
determine its nature because no evidence for softening could be found.
We note that there are other high data points which stand out in the ASM light
curve (e.g.\ near MJD\,50623: 1997 June 24), which might be remnants of other long X-ray flares. 
We examined the time-series data mode for each dwell that yielded an
anomalously high intensity for KS~1731$-$260 and found no evidence that a 
X-ray burst had occurred during any of these observations, which would 
increase the mean count rate during a dwell.
Although some of the high excursions occur when the Sun was close in position in the sky to \ks1731\ (e.g.,
near MJD\,50400: 1996 November 13), suggesting that the increase is not due to the
source itself, we can not rule out that the other high data points were obtained during 
long X-ray flares (if so, this would indicate recurrence times of less than a year).

\begin{figure}
 \resizebox{\hsize}{!}{\includegraphics[clip, bb=34 194 428 702]{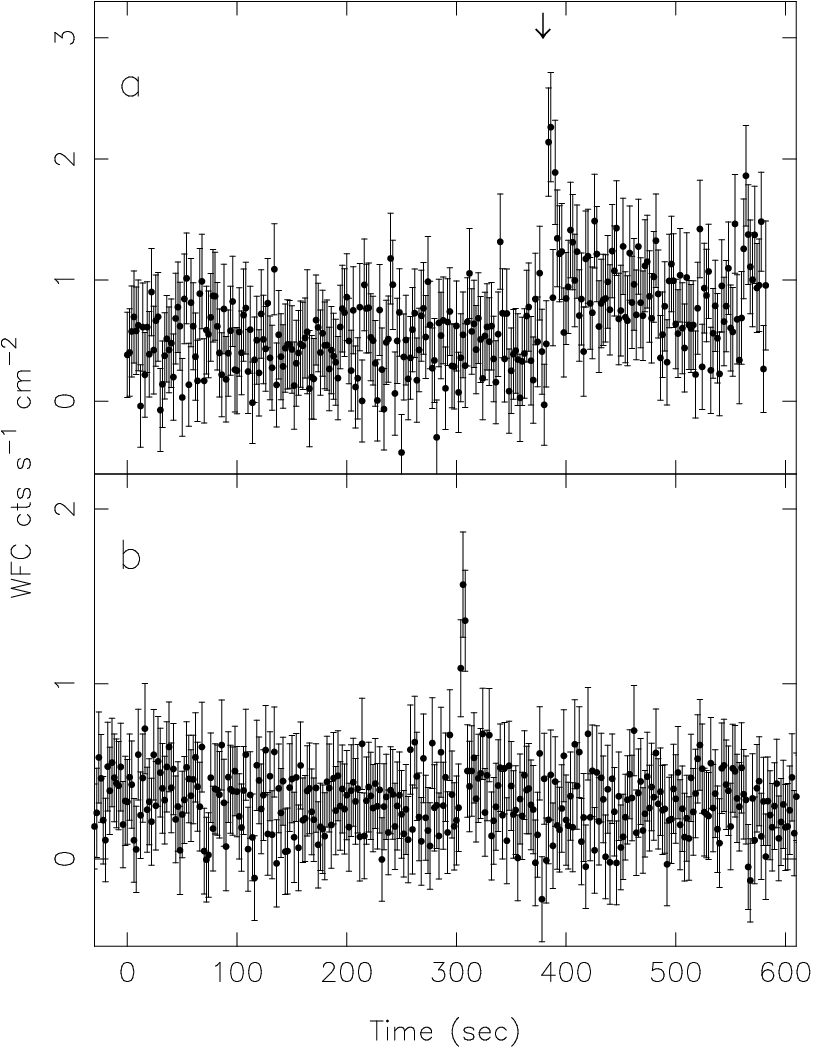}}
\caption{{\bf a.} WFC (2--28\,keV) count rates of the observation just before the long X-ray flare.
Time = 0\,s corresponds to UT 1996 September 23 09:52:02 (MJD\,50349.41113); 
the time resolution is 2\,s.
Clearly, a short and weak X-ray burst, which we call a precursor, can be seen 380\,s after the
start of the observation. The arrow points to the zero time scale of Fig.~3.
{\bf b.} Example of a WFC (2--28\,keV) observation around another short and weak X-ray burst.
Time = 0\,s corresponds to UT 1997 April 13 19:15:00 (MJD\,50551.80208); 
the time resolution is 2\,s.
}
\label{plot_precursor}
\end{figure}

\begin{figure}
 \resizebox{\hsize}{!}{\includegraphics[clip, bb=34 73 424 769]{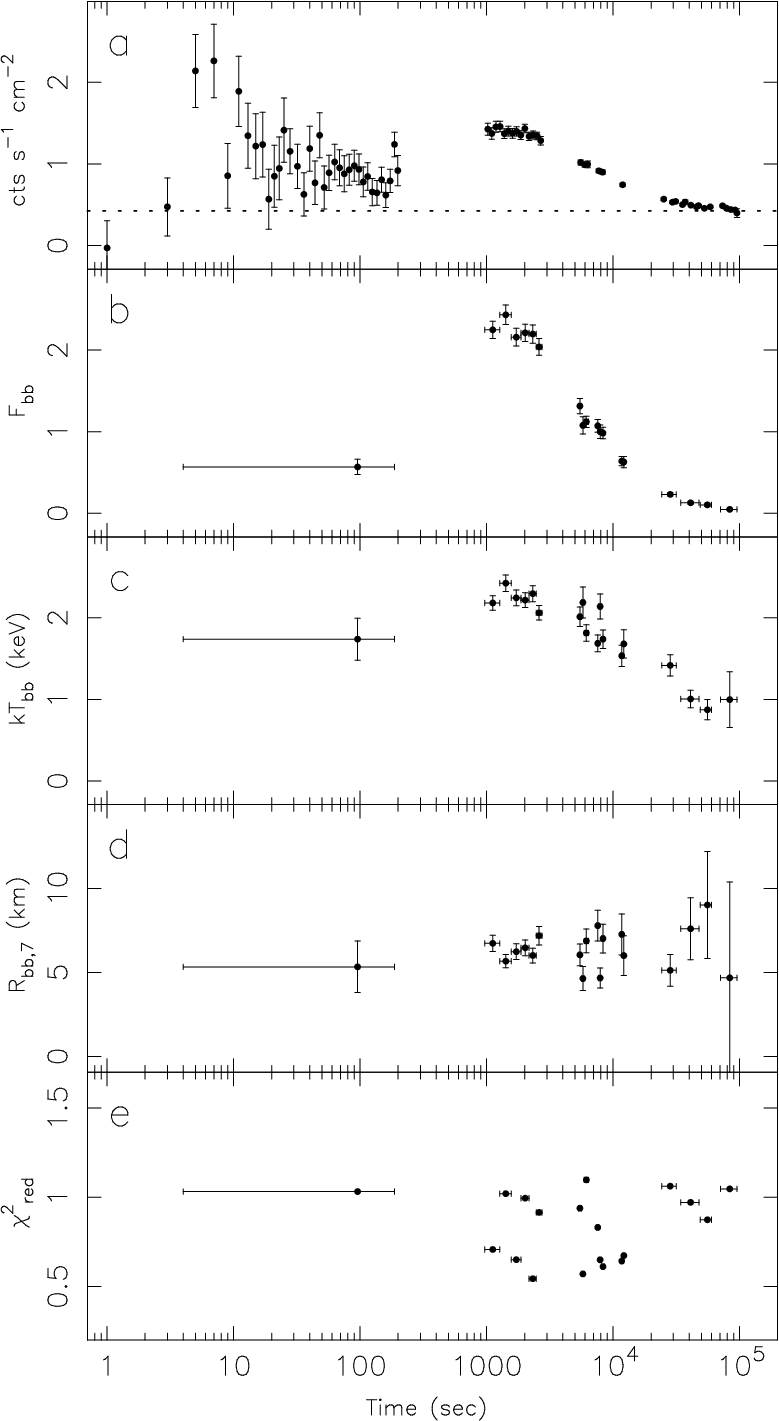}}
\caption{{\bf a.} WFC (2--28\,keV) count rates of the large flare
using a logarithmic time scale, with the data rebinned logarithmically
in time.  Time = 0\,s corresponds to UT 1996 September 23 09:58:21 (MJD\,50349.41552). 
The dotted line shows the mean count rate before the large X-ray
flare.  {\bf b--e.} Results of black-body fits to the net burst
emission (= total burst emission minus pre-burst emission): bolometric
black-body flux, $F_{\rm bb}$, in units of
10$^{-8}$\,erg\,cm$^{-2}$\,s$^{-1}$ ({\bf b}), black-body temperature,
$kT_{\rm bb}$ ({\bf c}), apparent black-body radius at 7\,kpc, $R_{\rm
bb,7}$ ({\bf d}), and the goodness of fit expressed in reduced
$\chi^2$ for 23 degrees of freedom ({\bf e}).  }
\label{plot_lc_pars_bb}
\end{figure}

In Fig.~\ref{plot_precursor}a, we show the count rates (2--28\,keV) of
the observation near the start of the X-ray flare.
The data show a short increase in intensity for about 4\,s, about 380\,s after the
start of the observation. We will refer to this as `a precursor'
hereafter. Afterwards, the intensity decreases somewhat erratically
to almost its pre-burst level\footnote{In Figs.~\ref{plot_asm_wfc}c
and d, the flux appears to increase steadily during the start of the
flare.  This is due to the choice of data rebinning.}.  Just before
the data gap, the flux may start to rise again, but unfortunately we
cannot be certain. In Fig.~\ref{plot_lc_pars_bb}a, we again show the count rates, 
now including the X-ray flare, on a logarithmic time scale.
After the gap ($\simeq$13\,min later), the flux is
high and it is possible the flare has already passed its maximum.

We fit the decay light curve of the precursor with an exponential
and derive the peak intensity, integrated count rate, and exponential
decay time, $\tau_{\rm exp}$.  In Table~1, we compare these numbers
with what is found for the other 63 X-ray bursts seen with the
WFC at comparable persistent pre-burst fluxes (i.e.\ before
MJD\,51000).  We find that the  characteristics of the precursor are comparable to the
other bursts, but note that the data are not of sufficient quality to
determine whether there occurs significant softening during its decay
(thus, we cannot prove it is a type~I burst). For comparison, we 
show a light curve around another similar short and weak X-ray burst (one of the 63 bursts)
in Fig.~\ref{plot_precursor}b, to illustrate that such bursts are not uncommon
in KS~1731$-$260 (see also Muno et al.\ 2000).

In Fig.~\ref{plot_asm_wfc}b, we show with tickmarks the times of
occurrence of normal X-ray bursts observed with the WFC around
the time of the long X-ray flare (no bursts were found during the short ASM
dwells on the \ks1731\ region in the same interval).  We found 20 normal
bursts, all of which occur before the flare.  The mean
burst rate, corrected for data gaps, is $\sim$3 per day, but the waiting times between bursts
are variable, the shortest being $\simeq$0.08\,day.  

{\em No} normal bursts were found during the WFC observations which were
obtained up to $\simeq$35\,days after the flare, despite
considerable exposure time (227\,ksec).  Given that 20 bursts occurred
during 589\,ksec of observation time before the flare we
would have expected to find $\simeq$8 bursts after the flare.  We
verified the significance of finding no bursts after the flare by
performing Monte Carlo simulations.  We randomly varied the burst
waiting times between the lowest observed waiting time (0.08\,day) and
0.6\,day (i.e.\ symmetric around 0.34\,day).  We then determined the
number of bursts which occurred within the observing windows after the
flare. By doing 10$^6$ simulations we found that the expected
number of bursts is indeed 7.6$\pm$2.4 (1\,$\sigma$); the probability of
observing no bursts after the flare is 1$\times$10$^{-4}$.  We
verified our method by applying the same procedure for the interval
before the flare: 20$\pm$3.5 are expected; indeed 20 are seen.

Another way to estimate the significance is to assume that the number
of observed normal bursts follows a Poisson distribution. In this case we
find that the probability of seeing no bursts instead of the 8 expected is
3.4$\times$10$^{-4}$.  Finally, in WFC observations during 1997 (first
observation was on MJD\,50509, i.e.\ 1997 March 2) and 1998, when the
source was at comparable flux levels, the source was also bursting,
again with similar burst frequencies as those before the flare
(see Fig.~\ref{plot_hardness}).

Given the above, we conclude that the burst activity was significantly
reduced or had ceased after the long X-ray flare.

\subsection{Spectral behaviour}

In order to derive the characteristics of the long X-ray flare emission, we
created spectra throughout its duration, including the
precursor. The time intervals were chosen such that we had sufficient
statistics per spectrum.  The net flare spectra (i.e.\ total source
spectra minus pre-flare persistent emission spectrum~C described
below) were satisfactorily fitted with emission from a black-body
($\chi^2_{\rm red}\lesssim$1 for 23 degrees of freedom).  
We note that some of the $\chi^2_{\rm red}$ values are rather small, 
perhaps due to a slight overestimate of the systematic errors. As we cannot
constrain interstellar absorption well with the WFC, we fixed $N_{\rm
H}$ to the value derived by Barret et al.\ (2000),
1.3$\times$10$^{22}$\,atoms\,cm$^{-2}$.  The results are shown in
Figs.~\ref{plot_lc_pars_bb}b--e.  The black-body temperature reaches a
maximum of $kT=2.4\pm0.1$\,keV and decreases thereafter.  The apparant
black-body radius at the distance of 7\,kpc remains more or less
constant throughout the burst with a value of 
$\simeq$6\,km; as expected, the relation 
between $kT$ and $F_{\rm bb}$ is consistent with a power law,
$F_{\rm bb}$$\propto$$T^{\alpha}$, with $\alpha=3.9$$\pm$0.4.
Similar black-body parameters are found for normal X-ray bursts from
this source (e.g.\ Muno et al.\ 2000).  

We also performed spectral fits to the flare spectra not subtracting 
the pre-flare persistent emission. The fits to the spectra
up to 12200\,s into the flare were acceptable: $\chi^2_{\rm red}=0.7-1.2$
(maximum $F_{\rm bb}=2.9\pm0.1\times 10^{-8}{\rm\,erg\,cm^{-2}\,s^{-1}}$, 
maximum $kT=2.2\pm0.1$\,keV, $R_{\rm bb,7}\simeq 10$\,km).
However, the fits to the spectra of the remainder of the burst 
were not acceptable: $\chi^2_{\rm red}=2.8-7.7$. This indicates that
while it seems more likely the persistent emission was present
during the flare, we can not rule out its absence during the 
first $\sim$3.5\,hr. In the remainder of the paper
we use the result from the spectral fits to the net-flare emission.

We estimated the fluence, $E_{\rm b}$, of the flare by assuming
that the rise from the start to the observed maximum is linear and
that the decay after this maximum is described by an
exponential.  In Table~1, we list the values of $E_{\rm b}$, the
maximum observed net-burst black-body flux, $F_{\rm bb,max}$, the
ratio $\tau\equiv E_{\rm b}/F_{\rm bb,max}$, and the ratio
$\gamma\equiv F_{\rm pers}/F_{\rm bb,max}$ (where F$_{\rm pers}$ is
the bolometric persistent source flux approximated by extrapolating the
observed persistent spectrum between 0.01 and 100\,keV).
Note that the values of $\tau$ and $\gamma$ are rather uncertain since we 
missed the peak of the flare.
We find that the fluence $E_{\rm b}$ of the flare
is a factor of 250--425 larger than what is observed for normal 
X-ray bursts in this source (see e.g.\ Muno et al.\ 2000).  
We note that if we do not use the exponential but simply 
integrate the observed fluxes during the flare and
interpolate over gaps, we obtain a slightly higher fluence of $\simeq 2.4\times 10^{-4}{\rm\,erg}$.
The flare does not reach the Eddington limit during our
observations: $F_{\rm bb,max}\simeq0.4\,L_{\rm Edd}$ given the
Eddington flux of $6.3\times10^{-8}{\rm\,erg\,cm^{-2}\,s^{-1}}$
inferred by Muno et al.\ (2000) from radius expansion bursts of
\ks1731.

\begin{table*}
\caption{\ks1731\ precursor and superburst characteristics}
\begin{tabular}{ccc}
\hline
 & {\bf Precursor} & {\bf Type I bursts} \\
peak rate$^a$ & 2.3$\pm$0.4\,cts\,s$^{-1}$\,cm$^{-2}$ & 1.8--5.3 (3.6) cts\,s$^{-1}$\,cm$^{-2}$ \\
integrated & & \\
count rate & $\simeq$13\,cts\,cm$^{-2}$ & 13--45 (29) cts\,cm$^{-2}$\\
$\tau_{\rm exp}$ (2--28\,keV) & 5.6$^{+2.8}_{-1.9}$\,s & 2--11 (5.5) s \\
\hline
\multicolumn{3}{c}{~} \\
{\bf Superburst} & & \\
\hline
parameter & value & $\chi^2_{\rm red}$/dof \\
$\tau_{\rm exp}$ (2--28\,keV) & 2.7$\pm$0.1\,hr & 0.94/116 \\
 (2--5\,keV)  & 4.4$\pm$0.3\,hr & 0.7/116 \\
 (5--28\,keV) & 2.1$\pm$0.1\,hr & 0.8/116 \\ 
\hline
parameter & at earth & at 7\,kpc \\
F$_{\rm pers}$$^b$ & 1.0$\pm$0.1$\times$10$^{-8}$\,erg\,s$^{-1}$\,cm$^{-2}$ & 5.9$\pm$0.8$\times$10$^{37}$\,erg\,s$^{-1}$ \\
E$_{\rm b}$ & $\simeq$1.7$\times$10$^{-4}$\,erg\,cm$^{-2}$ & $\simeq$9.8$\times$10$^{41}$\,erg\\
F$_{\rm bb,max}$ & 2.4$\pm$0.1$\times$10$^{-8}$\,erg\,s$^{-1}$\,cm$^{-2}$ & 1.4$\pm$0.1$\times$10$^{38}$\,erg\,s$^{-1}$ \\
\hline
parameter & value & \\
$\tau$=E$_{\rm b}$/F$_{\rm bb,max}$ & $\simeq$2.0\,hr & \\
$\gamma$=F$_{\rm pers}$/F$_{\rm bb,max}$ & 0.41$\pm$0.06 & \\
\hline
\multicolumn{3}{l}{\footnotesize $^a$\,For the normal bursts we give the range, and the average value in parentheses.} \\
\multicolumn{3}{l}{\footnotesize $^b$\,Unabsorbed pre-superburst luminosity (0.01--100\,keV).} \\
\end{tabular}
\end{table*}

\subsection{Persistent emission}

\begin{figure}
 \resizebox{\hsize}{!}{\includegraphics[clip, bb=38 496 421 769]{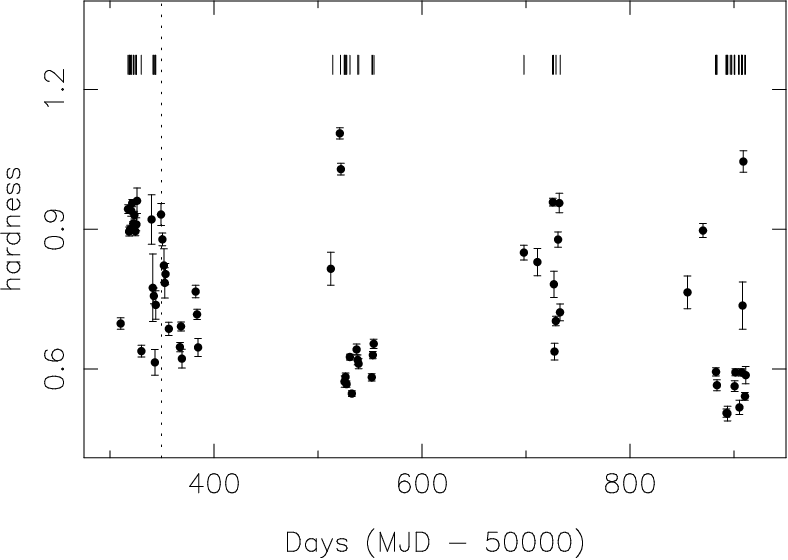}}
\caption{WFC hardness (as defined in Fig.~1e) curve from MJD\,50275--50950
(1996 July 11 -- 1998 May 17) when the persistent X-ray intensity
was at comparable levels.
Data points are 1-day averages. The dotted line denotes the time of the long X-ray flare.
In the upper part we have marked the times of the bursts with bars. Clearly,
no bursts are found just after the flare.
Normal bursting had resumed in the next observation.
}
\label{plot_hardness}
\end{figure}

The change in bursting behaviour just after the flare led us to
investigate whether the persistent emission is influenced as well.
We calculated hardness ratios (as defined in
Sect.~3.1) during the interval 1996--1998, when the source was at
comparable persistent flux levels.  These are shown in
Fig.~\ref{plot_hardness}; clearly, the source did not have 
exceptional hardness values after the flare.

In order to quantify the above, we created a total of six X-ray
spectra of the persistent emission before and after the 
flare during different intervals between MJD\,50308--50388 (see
Table~2).  For each spectrum data from various contiguous observations
with the same pointing were accumulated.  To parametrize the
persistent emission we fitted the 2--20\,keV spectra to an absorbed
cut-off power-law model plus a gaussian line at 6.4\,keV (see Barret
et al.\ 2000).  As above, we fixed the value of N$_{\rm H}$ to
1.3$\times$10$^{22}$\,atoms\,cm$^{-2}$; furthermore, we fixed the line
width of the gaussian, $\sigma$, to 0.1. This provided a good
description of the spectra\footnote{Previously, a black-body component
(or a disk black-body component) has been included in spectral fits
(Barret et al.\ 2000; Narita et al.\ 2001). We, however, found their
inclusion made the fits rather unstable, while providing little
(spectra A and F) or no (B--E) improvement.}; the results are
presented in Table~2.  The unabsorbed persistent luminosity before and
after the flare was $\simeq0.1\,L_{\rm Edd}$ (2--10\,keV).

\begin{table*}
\caption{Persistent emission spectral fit (absorbed cut-off power-law) parameters$^a$}
\begin{tabular}{lcccccc}
\hline
 & {\bf A} & {\bf B} & {\bf C} & {\bf D} & {\bf E} & {\bf F} \\
Interval (MJD$-$50000) & 318.6--322.2 & 323.4--326.0 & 348.7--349.3 & 350.2--250.5 & 366.7--369.0 & 383.1--385.0 \\
$F_{\rm 2-10\,keV}$ (10$^{-9}$\,erg\,s$^{-1}$\,cm$^{-2}$) & 4.6$\pm$0.2 & 4.8$\pm$0.3 & 5.2$\pm$1.1 & 5.7$\pm$1.4 & 4.3$\pm$0.4 & 4.2$\pm$0.8 \\
$\Gamma$$^b$ & 1.04$\pm$0.03 & 1.20$\pm$0.05 & 1.0$\pm$0.1 & 1.2$\pm$0.2 & 1.03$\pm$0.06 & 0.9$\pm$0.1 \\
$E_{\rm cut}$ (keV)$^c$ & 4.3$\pm$0.1 & 4.6$\pm$0.2 & 5.2$\pm$0.7 & 5.6$\pm$1.0 & 4.5$\pm$0.2 & 3.4$\pm$0.3 \\
norm.$^d$ & 1.34$\pm$0.03 & 1.63$\pm$0.05 & 1.3$\pm$0.1 & 1.5$\pm$0.2 & 1.17$\pm$0.05 & 1.26$\pm$0.09 \\
gnorm.$^e$ ($\times$10$^{-3}$) & 4.1$\pm$0.8 & 6.6$\pm$1.3 & ---$^f$ & ---$^f$ & ---$^f$ & ---$^f$ \\
$\chi^2$/dof & 1.6/21 & 0.9/21 & 1.2/22 & 1.1/22 & 1.0/22 & 1.5/22 \\
\hline
\multicolumn{7}{l}{\footnotesize $^a$\,N$_{\rm H}$ was fixed to 1.3$\times$10$^{22}$\,atoms\,cm$^{-2}$, see text.} \\
\multicolumn{7}{l}{\footnotesize $^b$\,Power-law photon index.} \\
\multicolumn{7}{l}{\footnotesize $^c$\,Cut-off energy.} \\
\multicolumn{7}{l}{\footnotesize $^d$\,Power-law nomalisation in photons\,cm$^{-2}$\,s$^{-1}$ at 1\,keV.} \\
\multicolumn{7}{l}{\footnotesize $^e$\,Normalisation in photons\,cm$^{-2}$\,s$^{-1}$ in the line modeled by a Gaussian fixed at 6.4\,keV and width ($\sigma$) of 0.1\,keV.} \\
\multicolumn{7}{l}{\footnotesize $^f$\,Including a gaussian did not improve the fit significantly.} \\
\end{tabular}
\end{table*}

We conclude that the persistent X-ray spectra days to weeks after the flare
(when no normal bursts occurred) are not different from the spectra days
to weeks before and months to years after the flare (when
normal bursts did occur).

\section{Discussion}

\subsection{Very long thermonuclear runaway events}

In many of its characteristics, the long X-ray flare resembles 
type~I bursts from \ks1731\ (e.g.\ Muno et al.\ 2000) as well as
type~I bursts from other sources.  The rise to maximum is relatively fast with
respect to the decay, the decay is exponential, and the spectra soften
(indicating cooling) during the decay.  This strongly suggests a
thermo-nuclear runaway event on the neutron star.  The duration of the
X-ray flare, however, is about half a day, i.e., a factor of
$\sim$1500--4000 longer than normal X-ray bursts from this
source.  Therefore, the fluence, $\simeq\!10^{42}$\,erg (at 7\,kpc),
is also much larger than typical.  The longest `normal' X-ray bursts seen 
in general are much shorter and less energetic, and all are
radius expansion events: they had durations on the
order of half an hour (Hoffman et al.\ 1987: probably from
4U~1708$-$23; Kuulkers et al.\ 2001: GX~17+2), fluences of
$\simeq$10$^{41}$\,erg, exponential decay times of 5.5\,min
(4U~1708$-$23) and 4.6\,min (GX~17+2), and ratios of the fluence to
peak net-burst flux ($\tau$) of $\simeq$5.4\,min and 6.6\,min,
respectively. These values are an order of magnitude smaller than those for
the superbursts.

\begin{table*}
\caption{Properties of superbursts ordered along their exponential decay time$^a$}
\begin{tabular}{ccccccc}
\hline
source               & {\bf 4U~1820$-$30} & {\bf Ser~X-1} & {\bf 4U~1735$-$44} & {\bf 4U~1636$-$53} & {\bf GX~3+1}  & {\bf KS~1731$-$260} \\
instrument           & PCA      & WFC & WFC      & ASM,PCA  & ASM & WFC,ASM \\
energy range         & 2--60\,keV & 2--28\,keV & 2--28\,keV & 1.5--12\,keV & 1.5--12\,keV & 2--28\,keV \\
duration (hr)        & $>$2.5 & $\sim$4 & $\sim$7 & $\gtrsim$1--3 & $>$3.3 & $\sim$12 \\
precursor?           & yes & ? & ? & ? & ? & yes \\
$\tau_{\rm exp}$ (hr) & $\simeq$1 & 1.2$\pm$0.1 & 1.4$\pm$0.1 & 1.5$\pm$0.1/3.1$\pm$0.5 & 1.6$\pm$0.2 & 2.7$\pm$0.1 \\
$L_{\rm pers}$ (L$_{\rm Edd}$)$^b$ & $\simeq$0.1 & $\simeq$0.2 & $\simeq$0.25 & $\sim$0.1 & $\sim$0.2 & $\simeq$0.1 \\
$kT_{\rm max}$ (keV) & $\simeq$3.0 & 2.6$\pm$0.2 & $\simeq$2.6$\pm$0.2 & ? & $\sim$2 & 2.4$\pm$0.1 \\
$L_{\rm peak}$ (10$^{38}$\,erg\,s)$^{c,d}$ & 3.4$\pm$0.1 & 1.6$\pm$0.2 & 1.5$\pm$0.1 & $\sim$1.2 & $\sim$0.8 & 1.4$\pm$0.1 \\
$E_{\rm b}$ (10$^{42}$\,erg)  & $\gtrsim$1.4 & $\simeq$0.8 & $\gtrsim$0.5 & $\sim$0.5--1 & $\gtrsim$0.6 & $\simeq$1.0 \\
$\tau$$\equiv$E$_{\rm b}$/L$_{\rm peak}$ (hr)$^d$ & $\gtrsim$1.1 & $\simeq$1.4 & $\gtrsim$0.9 & $\sim$1.2--2.3 & $\gtrsim$2.1 & $\simeq$2.0 \\
$\gamma$$\equiv$L$_{\rm pers}$/L$_{\rm peak}$$^d$ & $\simeq$0.1 & $\simeq$0.4 & $\sim$0.4 & $\sim$0.2 & $\sim$0.5 & $\simeq$0.4 \\
$t_{\rm no\,\,bursts}$ (days)$^e$ & ? & $\sim$34 & $>$7.5 & ? & ? & $>$35 \\
H/He or He donor     & He & ? & H/He & H/He & ? & ? \\
references$^f$       & S00,SB01 & C01 & C00 & W01 & K01 & this paper \\
\hline
\multicolumn{7}{l}{\footnotesize $^a$\,A question mark denotes an unknown value.} \\ 
\multicolumn{7}{l}{\footnotesize $^b$\,We used the 0.01--100\,keV unabsorbed flux from spectral fits and the observed maximum flux} \\ 
\multicolumn{7}{l}{\footnotesize $^{~}$\,during radius expansion bursts.} \\
\multicolumn{7}{l}{\footnotesize $^c$\,Unabsorbed bolometric peak (black-body) luminosity.} \\
\multicolumn{7}{l}{\footnotesize $^d$\,Since the rise to maximum was only observed for 4U\,1820$-$30, the values for the other sources are to be used with caution.} \\
\multicolumn{7}{l}{\footnotesize $^e$\,Time of cessation of normal X-ray bursts after the superburst.} \\
\multicolumn{7}{l}{\footnotesize $^f$\,C00=Cornelisse et al.\ 2000, C01=Cornelisse et al.\ 2001, K01=Kuulkers 2001, S00=Strohmayer 2000,} \\
\multicolumn{7}{l}{\footnotesize $^{~}$\,S01=Strohmayer \&\ Brown 2001, W01=Wijnands 2001.} \\
\end{tabular}
\end{table*}

The long X-ray flare resembles the superbursts seen in five other sources
(4U~1820$-$30, Ser~X-1, 4U~1735$-$44, 4U~1636$-$53 and GX~3+1), as can be
seen in Table~3, where we compare the properties of all seven 
flares observed to date. 
The duration of the superbursts are estimated by eye from the light curves 
and the hardness curves.
With respect to duration of about half a day the flare in \ks1731\ is the longest 
among the superbursts seen so far.  

The superburst seen in 4U~1820$-30$ (Strohmayer 2000; Strohmayer \&\ Brown 2001) 
is the only one
for which the full rise to maximum has been observed. That superburst
was immediately preceeded by a normal X-ray burst.  This
`precursor' lasted for $\simeq$20\,s.  Both the `precursor' and the
superburst exhibited radius expansion, indicating the luminosity
reached the Eddington limit. The superbursts from all sources but 4U~1820$-$30 
seem to be a factor 2--4 fainter than their brightest normal bursts, although we
should stress again that only in 4U~1820$-$30 the rise to maximum was
fully covered.  Next best is \ks1731, for which we observed the start
and found a short ($\simeq$4\,s) and weak burst, most probably a
type~I burst, preceding the superburst by $\gtrsim$200\,s.  
Note that the precursor
does not necessarily need to be a trigger of the superburst (see also Strohmayer \&\ Brown 2001).
The observation of precursors close to the superbursts is another difference
from normal (type~I) bursts, which do not have such precursors (see e.g.\ Lewin et
al.\ 1993). 

We found evidence that the occurrence of normal bursts stopped
for more than a month after the superburst in KS~1731$-$260. Similarly, no bursts
were observed for about a week after the superburst in
4U~1735$-$44, whereas in Ser~X-1 no bursts
were seen for $\sim$5 weeks after the superburst (Cornelisse et al.\ 2000, 2001; see also 
Table 3). This suggests that the 
superburst has profound influence on the outer envelope of the neutron star
where normal bursts originate.  The outer envelope may be
heated by the energy generated by superburst, causing the burning
of hydrogen and/or helium to become stable (see Cumming \&\
Bildsten 2001). However, 
the persistent emission of \ks1731\ during
the non-bursting interval is not significantly different compared to
intervals of normal bursting. This seems evidence against the idea that
the neutron star is significantly hotter, but might also just be because 
the observed persistent emission is dominated by emission from the accretion disk.  

\subsection{Unstable carbon burning...}

Assuming unstable nuclear processes, we can use the time scales and
total energy to estimate where the burning occurs, how much matter
needs to be involved, etc.\ (see also Cumming \&\ Bildsten 2001;
Strohmayer \&\ Brown 2001).  Since the fusion of elements higher than
hydrogen releases about 1\,MeV per nucleon
($\simeq10^{18}{\rm\,erg\,g^{-1}}$; see, e.g., Lamb \&\ Lamb 1978),
one needs to process roughly 10$^{24}$\,g of matter to produce the
total energy of 10$^{42}$\,erg released during a superburst.  With 
the mass accretion rates of $\sim$$10^{-9}\,{\rm M}_{\odot}{\rm\,yr^{-1}}$
(as appropriate for sources which radiate at $\sim\!0.1$--$0.3\,L_{\rm
Edd}$) the expected recurrence time is at least half a year (depending
on how much of the accreted matter is consumed during a
superburst).  This indicates that they are indeed rare events.  

The above mass estimate corresponds to an ignition column depth
$y=\Delta M/4\pi R_{\rm ns}^2\simeq8\times10^{10}{\rm\,g\,cm^{-2}}$ (where
$\Delta M$ is the amount of burned matter and $R_{\rm ns}$ the radius of the
neutron star; we take $R_{\rm ns}=10$\,km).  At these column depths, the
thermal time is on the order of 1.5--3\,hrs, almost independent 
of the temperature (see e.g.\ Brown
\&\ Bildsten 1998).  This is in excellent agreement with the decay
time scale of the superbursts. We note that the density at
column depths $\sim8\times10^{10}{\rm\,g\,cm^{-2}}$ is
$\rho\simeq10^8{\rm\,g\,cm^{-3}}$ (see, e.g., Brown \&\ Bildsten 
1998).  

The above arguments have led to the suggestion that superbursts are
due to unstable carbon burning deeper in the neutron star, in a layer
beneath the (un)stable hydrogen/helium or helium layer (Cumming \&
Bildsten 2001; Strohmayer \& Brown 2001).  Obviously, in order for
this to work, the ashes of the burning hydrogen/helium layer need to
contain carbon.  One way to achieve this is to have stable burning of
helium.  This may occur in 4U\,1820$-$30, in which the neutron star is
thought to accrete pure helium because the companion is most likely a helium white dwarf
(based on the very short orbital period (Rappaport et al.\ 1987),
which is consistent with the short bursts this source shows.
4U\,1820$-$30 shows long intervals in which no normal X-ray bursts occur, 
which gave ample time
to generate the necessary amount of carbon (Strohmayer \&\ Brown
2001).  

The above mechanism cannot work for all systems, however.  In
particular, for 4U~1735$-$44 and 4U~1636$-$53 it has been shown that
the donors provide hydrogen-rich material (Augusteijn et al.\ 1998).
For burning of a hydrogen/helium mix, one expects only small amounts
of carbon to be produced, both for the case that the burning is stable
and for the case that it is unstable (e.g.\ Schatz et al.\ 1999,
2001).  In particular, during normal bursts temperatures in excess of $\sim10^9$\,K
are reached and a break-out of the CNO cycle occurs, creating elements
far beyond the iron group.  Nevertheless, over a sufficiently long
time (many years for mass accretion rates of a few tenth of Eddington)
sufficient carbon may be produced to cause a superburst
(Cumming \&\ Bildsten 2001).  

Another issue is whether the burning of pure carbon can be unstable
at such low column depths as 8$\times$10$^{10}$\,g\,cm$^{-2}$. Brown \&\ Bildsten
(1998) found ignition at $>$5$\times$10$^{12}$\,g\,cm$^{-2}$
for pure carbon at Eddington accretion rates. The superbursters
accrete near a tenth of the Eddington accretion rate, making it even more of a challenge
to ignite carbon at the lower pressures implied by the observations.
Strohmayer \&\ Brown (2001) show that one may underestimate the 
amount of energy released when integrating the superburst X-ray flux, and 
therefore also the amount of carbon consumed during the superburst.
According to the authors, most of the energy 
is released as neutrinos, and most of the heat produced by the burning largely flows
inwards. If so, the large amounts of carbon required ($\gtrsim$10$^{26}$\,g), 
imply very long recurrence times for the superbursts in 4U\,1820$-$40
(decades).
An alternative suggestion is that the precursor helium flash triggered unstable
carbon burning, but Strohmayer \&\ Brown (2001) consider this to be 
less likely.

Cumming \&\ Bildsten (2001) have a different view. They point out that the heavy
elements in the ashes will reduce the thermal conductivity, which may lead to large 
temperature gradients and ignition at the observed column depth.
They find this works for mass accretion rates 
above 10\%\ of the Eddington rate and for mass fractions of carbon 
larger than 0.1. They argue that the fact that up to now
superbursts have only been reported in systems with 
accretion rates between 10 and 30\%\ of Eddington, is that 
at lower mass accretion rates the superbursts are easier to
distinguish from the persistent emission.

Cumming \&\ Bildsten (2001; see also Strohmayer \&\ Brown 2001) 
argued that the cooling following the carbon burning 
may provide a large heat flux into the hydrogen/helium burning layers,
stabilizing the burning there. For mass accretion rates of 10--20\%\
of Eddington bursting would then cease for 
about 15--20 days after the superburst. This is more or less consistent with what is observed in
KS\,1731$-$260 and Ser\,X-1. 

\subsection{... or runaway electron capture?}

Although with unstable carbon burning it appears possible to
understand the superbursts, there may be an interesting alternative.
The densities at which the Fermi energy of the electrons, which
provide the pressure in the outer layers, becomes equal to the
difference in rest-mass of a neutron and a proton are
2--3$\times$10$^7$\,g\,cm$^{-3}$ (where the exact number depends on
the composition).  This density corresponds to a column density of
$y\simeq 2\times 10^{10}g_{14.3}^{-1}$\,g\,cm$^{-2}$, where $g_{14.3}$
is the surface gravity normalised to $10^{14.3}$\,cm\,s$^{-2}$,
appropriate for a neutron star mass $M_{\rm ns}=1.4$\,M$_{\odot}$ and $R_{\rm ns}=10$\,km
(e.g.\ Bildsten \&\ Cumming 1998).  This is not too different from the
values inferred above.  Hence, if any protons are present, they will
capture electrons via the $p(e^-,\nu\/)n$ reaction (Rosenbluth et al.\
1973; see also Taam et al.\ 1996; Bildsten \&\ Cumming 1998).  The
released neutrons are then captured by heavy nuclei (rather than
by protons, see also below) releasing $\simeq$7--8\,MeV per nucleon
(Bildsten \&\ Cumming 1998).  Depending on the temperature (which sets
the precise distribution of electron energies) and the thermal
conductivity, this process may be unstable: an increase in temperature
leads to a larger fraction of the electrons at high energies; if the
region does not cool sufficiently fast, this leads to more captures, a
further increase in temperature, etcetera.

There have been some studies of the above described mechanism, but only for the case
in which no heavy nuclei are present and the neutrons are captured
by protons.  This process is called `deep hydrogen burning' (Hameury
et al.\ 1982, 1983; Woosley \&\ Weaver 1984; Fushiki et al.\ 1992;
Taam et al.\ 1993, 1996).  This type of burning can become unstable
when the neutron star envelope has a low temperature
($\lesssim$10$^7$\,K) and/or when the abundance of CNO elements is
very low ($Z\lesssim0.001$).  The latter may not be as unrealistic as it appears
at first, since the original CNO abundance received from the donor may be
significantly reduced due to nuclear spallation processes in the
accretion shock (Bildsten et al.\ 1992).  In the models, long
($>$30\,min) X-ray bursts have been produced at moderate accretion
rates ($\sim$0.1\,\.M$_{\rm Edd}$), with total nuclear energy releases
of up to several times 10$^{41}$\,erg (Woosley \&\ Weaver 1984, their model 6;
Fushiki et al.\ 1992).
This is still an order of magnitude less than
observed in superbursts (but rather similar to the `long' X-ray bursts
seen from 4U~1708$-23$ and GX~17+2!).  

Several other questions have to be answered as well, however.
Foremost, whether there is (enough) hydrogen left after the
(un)stable hydrogen/helium burning.
Various calculations
suggest residual hydrogen mass fractions of $X_H\simeq$0.04--0.2 for
mass accretion rates of $>$10$^{-9}$\,M$_{\odot}$\,yr$^{-1}$ 
(e.g.\ Taam et al.\ 1996; see also Koike et al.\ 1999).
(Of course, this mechanism
cannot explain the superburst in 4U\,1820$-$30, since for that source
there are no protons to begin with.  We note, however, that this
superburst seems to be somewhat different from the others, so a
different mechanism may not be unexpected.)

To produce $10^{42}{\rm\,erg}$ one needs $\simeq$10$^{23}$\,g of
residual hydrogen (given the energy release of 7--8\,MeV per nucleon).
Thus a total accreted matter of 5--25$\times$10$^{23}$\,g is needed;
for accretion rates of $\sim$10$^{-9}$\,M$_{\odot}$\,yr$^{-1}$ the
recurrence times of superbursts would then be 0.25--1.25\,yr. This is
in an order of magnitude shorter than the case for unstable carbon
burning as estimated by Cumming \&\ Bildsten (2001).  We estimate the
maximum expected energy release for runaway electron capture from
$E_{\rm max}=4\pi R_{\rm ns}^2yE_{\rm nuc}X_H= 2\times10^{41}R_{\rm
ns,10}^4X_{H,0.1}\,$erg, where $R_{\rm ns}=10R_{\rm ns,10}\,$km is the
radius of the neutron star, $E_{\rm nuc}$ the above quoted energy
release, and $X_H=0.1X_{H,0.1}$ (the
fourth-power dependence on radius arises because of the dependence of
$y$ on the surface gravity).  This may be consistent with the observed
value of $10^{42}\,$erg, if the radius of the neutron star is larger
than 10\,km, as is indeed inferred for many equations of state (see,
e.g., the review by Lattimer \& Prakash 2000), and/or the residual mass
fraction of hydrogen is larger than~0.1.  Thus, unstable electron
capture appears a viable alternative mechanism for the superbursts.

\begin{acknowledgements}
We thank Lars Bildsten, Andrew Cumming and Ron Taam for discussions,
in particular about the possibility of the superbursts being due to
electron capture.  The BeppoSAX satellite is a joint Italian and Dutch
programme. We made use of quick-look results provided by the ASM/RXTE
team.  MHvK acknowledges support from the Royal Netherlands Academy of 
Science KNAW.
\end{acknowledgements}


\begin{thebibliography}{}

\bibitem[1998]{}
Augusteijn, T., van der Hooft, F., de Jong, J.A., van Kerkwijk, M.H., van Paradijs, J. 1998, A\&A, 332, 561
\bibitem[2000]{}
Barret, D., Olive, J.F., Boirin, L., Done, C., Skinner, G.K., Grindlay, J.E. 2000, ApJ, 533, 329
\bibitem[1992]{}
Bildsten, L., Salpeter, E.E., Wasserman, I. 1992, ApJ, 384, 183
\bibitem[1998]{}
Bildtsen, L., Cumming, A. 1998, ApJ, 506, 842
\bibitem[1997]{}
Boella G., Butler R.C., Perola G.C., et al., 1997, A\&AS 122, 299
\bibitem[2000]{}
Brown, E.F. 2000, ApJ, 531, 988
\bibitem[1998]{}
Brown, E.F., Bildsten, L. 1998, ApJ, 496, 915
\bibitem[2000]{}
Cornelisse, R., Heise, J., Kuulkers, E., Verbunt, F., in~'t~Zand, J.J.M. 2000, A\&A, 357, L21
\bibitem[2001]{}
Cornelisse, R., Kuulkers, E., in~'t~Zand, J.J.M., Verbunt, F., Heise, J. 2001, A\&A, submitted
\bibitem[2001]{}
Cumming, A., Bildsten, L. 2001, ApJ, 559, L127
\bibitem[1992]{}
Fushiki, I., Taam, R.E., Woosley, S.E., Lamb, D.Q. 1992, ApJ, 390, 634
\bibitem[1982]{}
Hameury, J.M., Bonazzola, S., Heyvaerts, J., Ventura, J. 1982, A\&A, 111, 242
\bibitem[1983]{}
Hameury, J.M., Heyvaerts, J., Bonazzola, S. 1983, A\&A, 121, 259
\bibitem[1978]{}
Hoffman, J.A., Lewin, W.H.G., Doty, J., Jernigan, J.G., Haney, M., Richardson, J.A. 1978,
ApJ, 221, L57
\bibitem[1997]{}
Jager, R., Mels, W.A., Brinkman, A.C., et al. 1997, A\&ASS, 125, 557
\bibitem[1999]{}
Koike, O., Hashimoto, M., Arai, K., Wanajo, S. 1999, A\&A, 342, 464
\bibitem[2001]{}
Kuulkers, E. 2001, ATel \#68
\bibitem[2000]{}
Kuulkers, E., in~'t~Zand, J.J.M., Cornelisse, R., et al. 2000, A\&A, 358, 993
\bibitem[2001]{}
Kuulkers, E., Homan, J., van der Klis, M., Lewin, W.H.G., M\'endez, M. 2001, 
A\&A, submitted [astro-ph/0105386]
\bibitem[1978]{}
Lamb, D.Q., Lamb, F.K. 1978, ApJ, 220, 291
\bibitem[2001]{}
Lattimer, J.M., Prakash, M. 2001, ApJ, 550, 426
\bibitem[1984]{}
Lewin, W.H.G., Vacca, W.D., Basinska, E.M. 1984, ApJ, 277, L57
\bibitem[1993]{}
Lewin, W.H.G., van Paradijs, J., Taam, R.E. 1993, Space Sci.\ Rev., 62, 223
\bibitem[2000]{}
Muno, M.P., Fox, D.W., Morgan, E.H., Bildsten, L. 2000, ApJ, 542, 1016
\bibitem[2001]{}
Narita, T., Grindlay, J.E., Barret, D. 2001, ApJ, 547, 420
\bibitem[1987]{}
Rappaport, S., Ma, C.P., Joss, P.C., Nelson, L.A. 1987, ApJ, 322, 842
\bibitem[1973]{}
Rosenbluth, M.N., Ruderman, M., Dyson, F., Bahcall, J.N., Shaham, J., Ostriker, J. 1973,
ApJ, 184, 907
\bibitem[2001]{}
Revnivtsev, M.G., Sunyaev, R.A. 2001, A\&A Letters, submitted [astro-ph/0108323]
\bibitem[2002]{}
Revnivtsev, M.G., Sunyaev, R.A. 2002, Astron.\ Lett., in press [astro-ph/0108120]
\bibitem[1999]{}
Schatz, H., Bildsten, L., Cumming, A., Wiescher, M. 1999, ApJ, 524, 1014
\bibitem[2001]{}
Schatz, H., Aprahamian, A., Barnard, V., et al. 2001, Phys.\ Rev.\ Lett., 86, 3471
\bibitem[1997]{}
Smith, D.A., Morgan, E.H., Bradt, H. 1997, ApJ, 479, L137
\bibitem[2001]{}
Smith, D.A., Levine, A., Bradt, H., Hurley, K., Feroci, M., Butterworth, P.,
Golenetskii, S., Pendleton, G. \&\ Phengchamnan, S. 2001, ApJS, in press
[astro-ph/0103357]
\bibitem[2000]{}
Strohmayer, T.E. 2000, HEAD 32, 24.10
\bibitem[2000]{}
Strohmayer, T.E., Brown, E.F. 2001, ApJ, in press [astro-ph/0108420]
\bibitem[1989]{}
Sunyaev, R.A., the Kvant Team 1989, IAU Circ.\ 4839
\bibitem[1990]{}
Sunyaev, R.A., Gilfanov, M., Churazov, E., et al. 1990, Sov.\ Astron.\ Lett., 16, 59
\bibitem[1993]{}
Taam, R.E., Woosley, S.E., Weaver, T.A., Lamb, D.Q. 1993, ApJ, 413, 324
\bibitem[1996]{}
Taam, R.E., Woosley, S.E., Lamb, D.Q. 1996, ApJ, 459, 271
\bibitem[1988]{}
van Paradijs, J., Penninx, W., Lewin W.H.G. 1988, MNRAS, 233, 437
\bibitem[2001]{}
Wijnands, R. 2001, ApJ, 554, L59
\bibitem[2001]{}
Wijnands, R., Miller, J.M., Markwardt, C., Lewin, W.H.G., van der Klis, M. 2001a, 
ApJ, 560, L159
\bibitem[2001]{}
Wijnands, R., Groot, P.J., Miller, J.L., Marwardt, C., 
Lewin, W.H.G., van der Klis, M. 2001b, ATel \#72
\bibitem[1997]{}
Wijnands, R.A.D., van der Klis M. 1997, ApJ, 482, L65
\bibitem[1984]{}
Woosley, S.E., Weaver, T.A. 1984, in High Energy Transients in 
Astrophysics, AIP, p.~275

\end{thebibliography}
\end{document}